\documentclass[12pt]{article}
\usepackage{graphicx}
\begin{document} 
\title{Phase transitions in Nowak-Sznajd opinion dynamics}

\author{Maciej Wo{\l}oszyn, Dietrich Stauffer* and Krzysztof Ku{\l}akowski\\
Faculty of Physics and Applied Computer Science\\
AGH University of Science and Technology\\
al. Mickiewicza 30, PL-30059 Krak\'ow, Euroland}

\maketitle
\noindent
* Visiting from Institute for Theoretical Physics, Cologne University,
D-50923 K\"oln, Euroland  

\bigskip
\noindent
e-mail: woloszyn@agh.edu.pl, stauffer@thp.uni-koeln.de, 

kulakowski@novell.ftj.agh.edu.pl

\bigskip
\small{
Abstract: The Nowak modification of the Sznajd opinion dynamics model on the
square lattice assumes that with probabilities $\beta$ and $\gamma$ the
opinions flip due to mass-media advertising 
from down to up, and vice versa. Besides, with probability 
$\alpha$ the Sznajd rule applies that a neighbour pair agreeing in its two 
opinions convinces all its six neighbours of that opinion. Our Monte Carlo
simulations and mean-field theory find sharp phase transitions in the 
parameter space.
}

\bigskip

\section{Introduction and Model}

\begin{figure}[hbt]
\begin{center}
\includegraphics[angle=-90,scale=0.3]{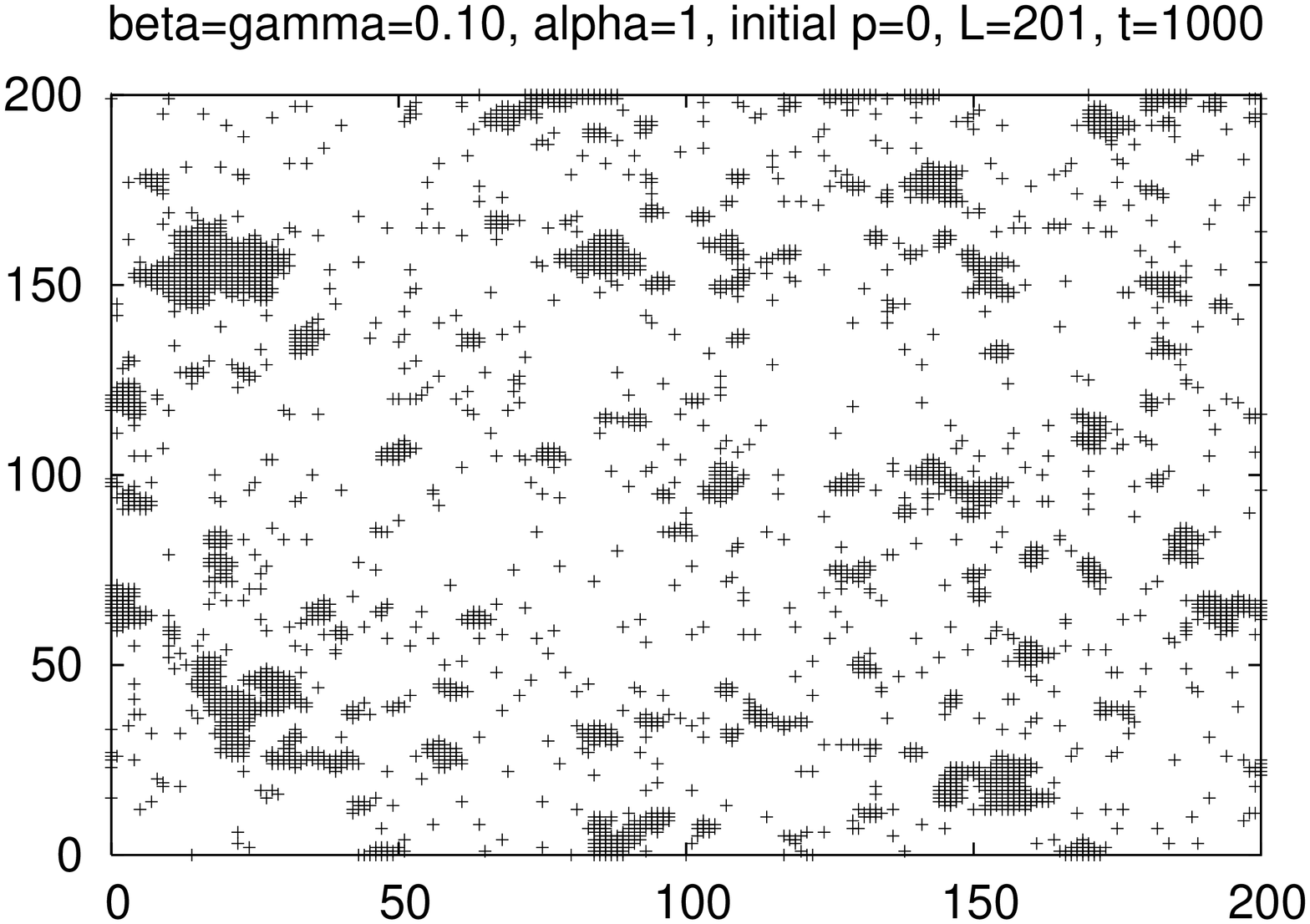}
\includegraphics[angle=-90,scale=0.3]{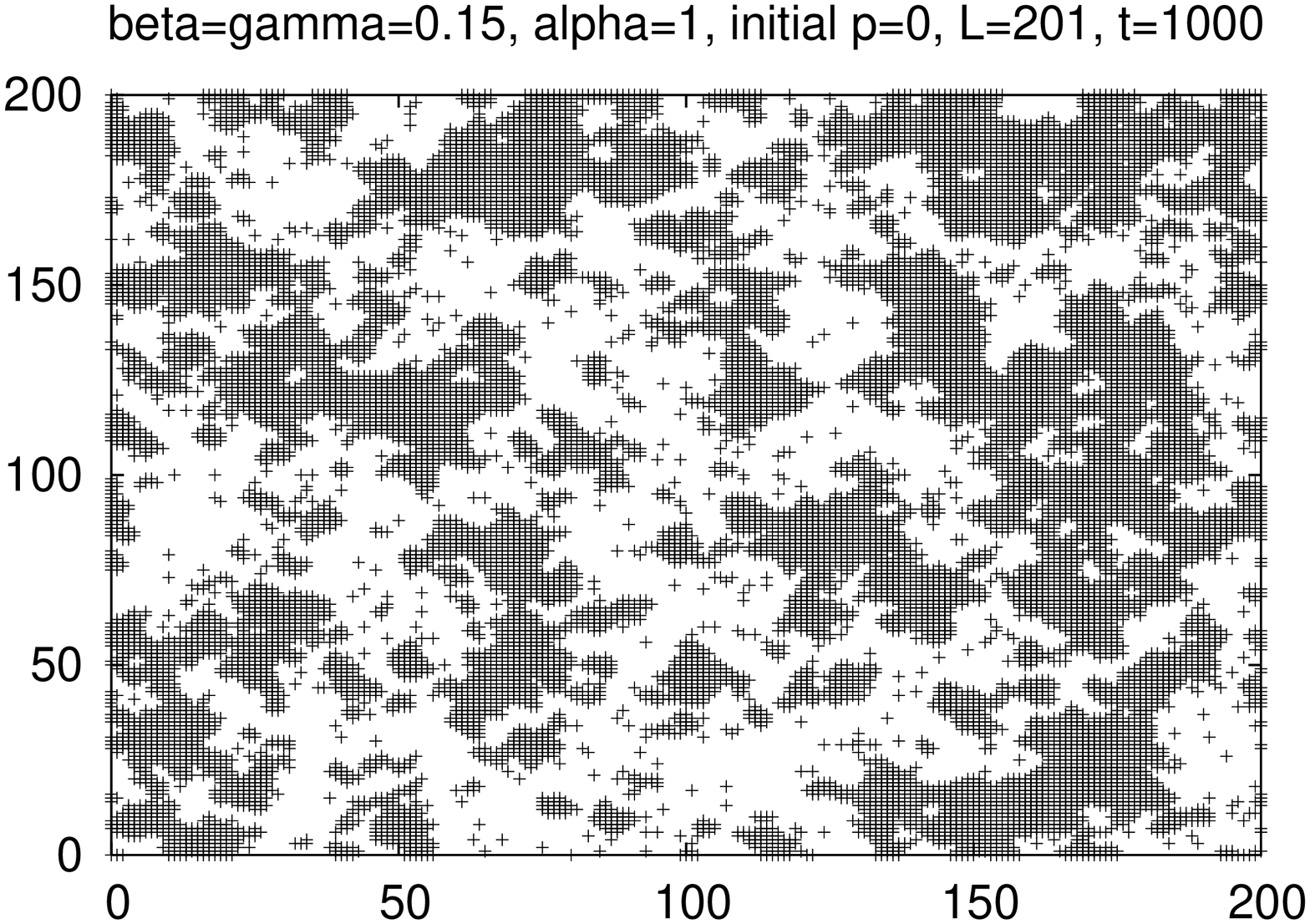}
\end{center}
\caption{Distribution of up opinions when initially one percent of the agents
had opinion up at randomly selected sites. The upper part had a lower flipping
probability $\beta = \gamma = 0.10$ than the lower part (0.15), and in the lower
part the two opinions occur about equally often.
}
\end{figure}

\begin{figure}[hbt]
\begin{center}
\includegraphics[angle=-90,scale=0.31]{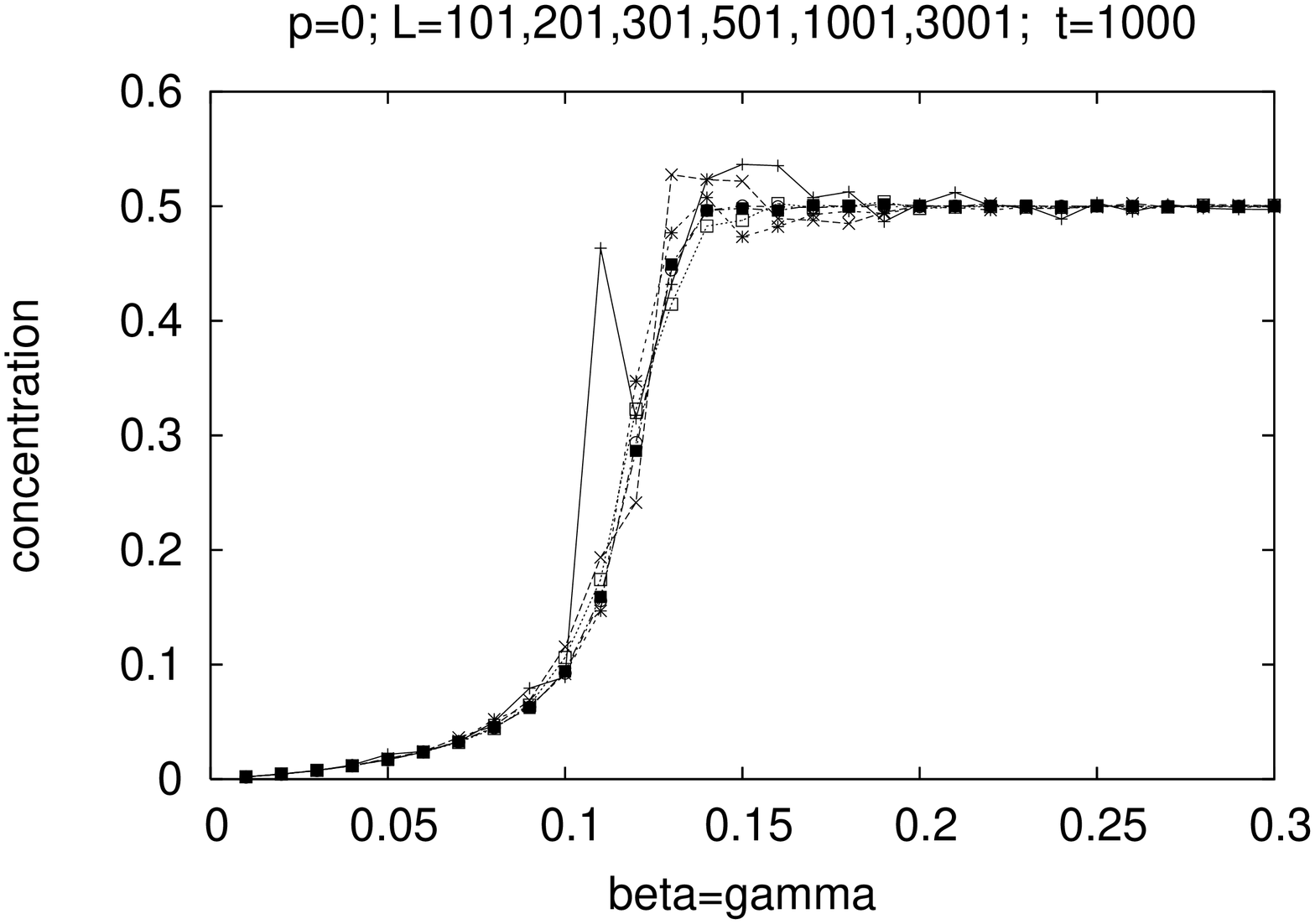}
\includegraphics[angle=-90,scale=0.31]{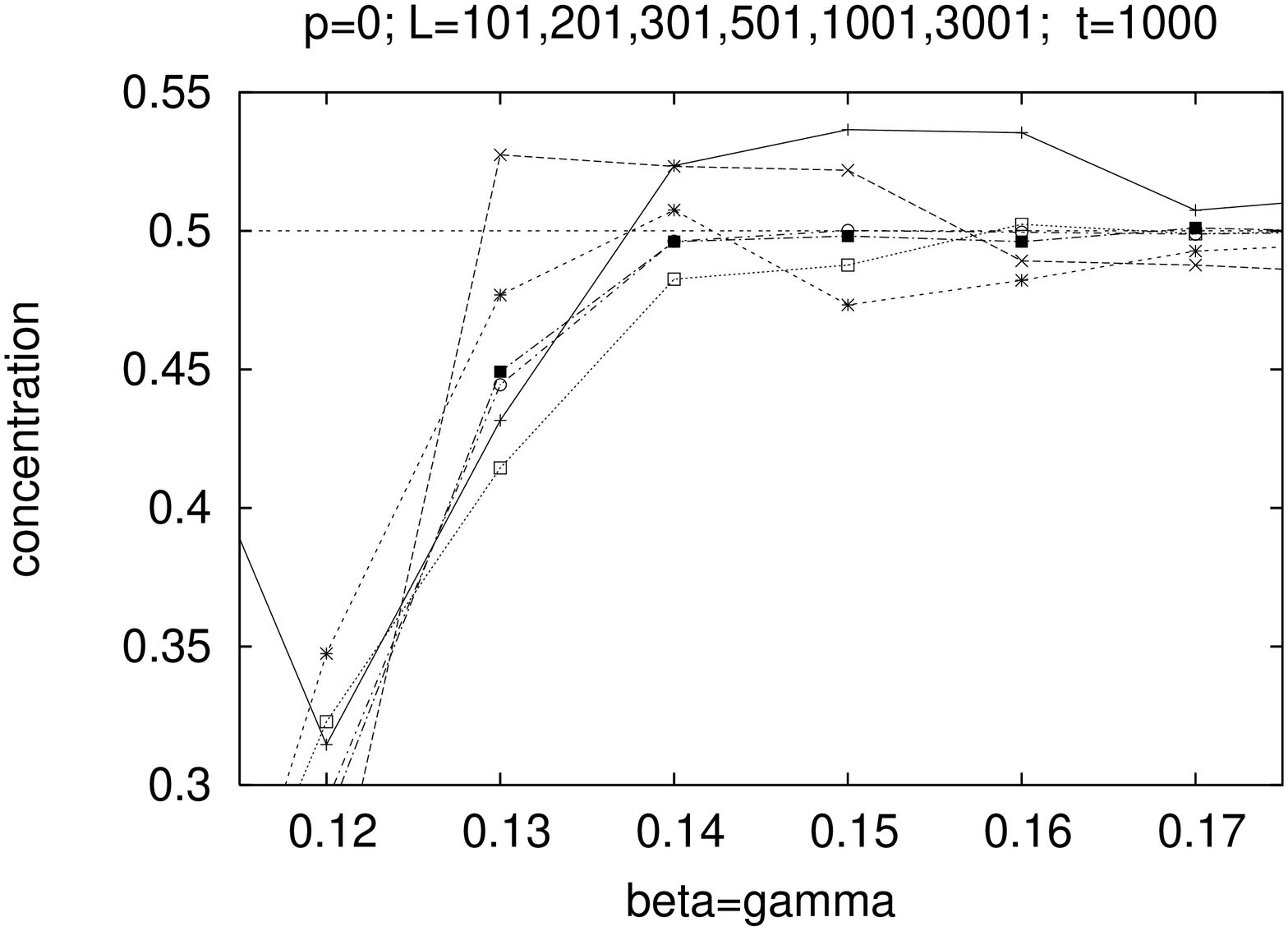}
\end{center}
\caption{Phase transition when starting from $p = 0.01$; one sample was 
simulated for each point, with $\alpha=1, \; t=1000, \; L$ = 101 (+), 201 (x),
301 (*), 501 (empty squares), 1001 (full squares) and 3001 (circles). 
}
\end{figure}

\begin{figure}[hbt]
\begin{center}
\includegraphics[angle=-90,scale=0.5]{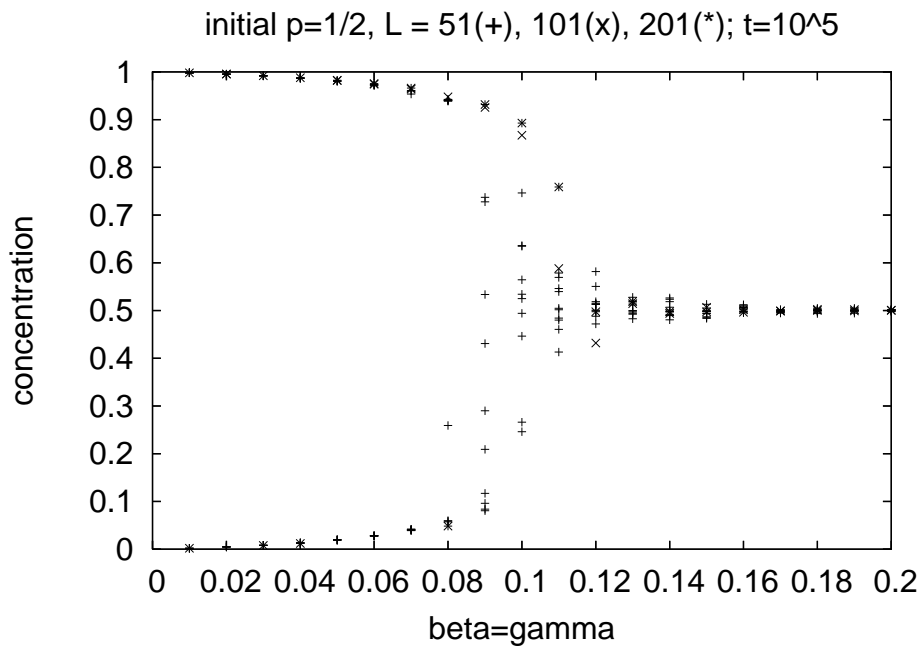}
\end{center}
\caption{As Fig.2 but starting from $p=0.5$; ten samples each for $L = 51$
and one for $L = 101$ 
and 201; $t = 10^5$. In some cases the larger lattices seem not
yet have found their equilibrium.
}
\end{figure}

\begin{figure}[hbt]
\begin{center}
\includegraphics[angle=-90,scale=0.5]{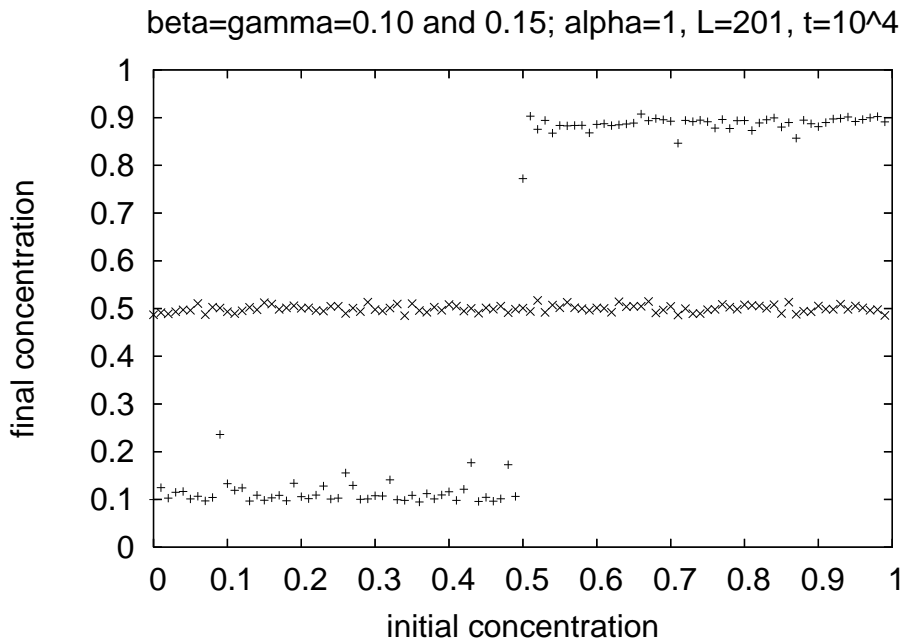}
\end{center}
\caption{Possible jump as a function of initial concentration if $\beta=\gamma$
is fixed at 0.10 (+, jump) and 0.15 (x, no jump).
}
\end{figure}

\begin{figure}[hbt]
\begin{center}
\includegraphics[angle=-90,scale=0.5]{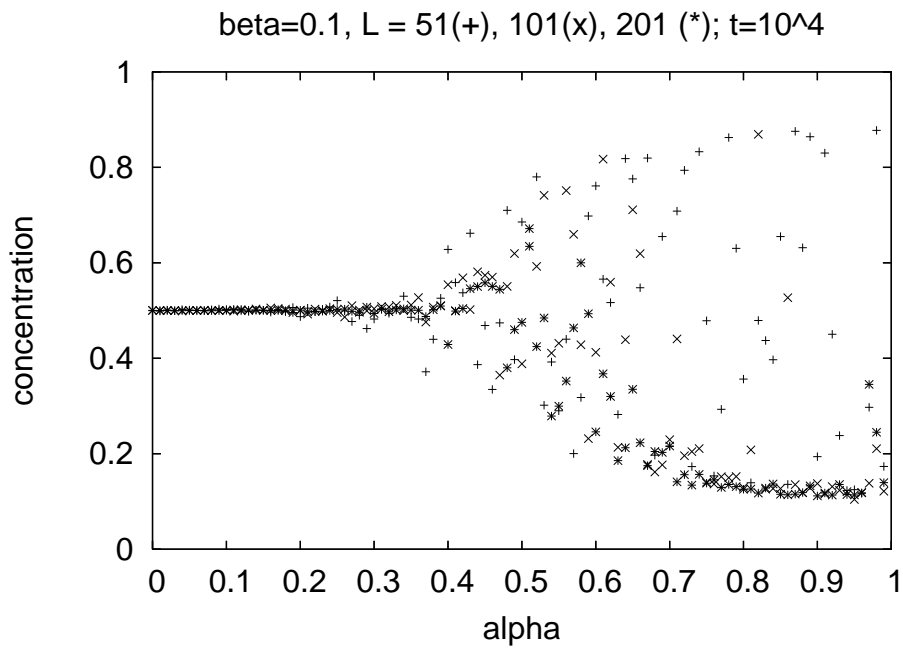}
\end{center}
\caption{Phase transition as a function of $\alpha$, at fixed $\beta=\gamma=0.1$. 
Only for $L > 100$ a broken symmetry is clearly visible for large $\alpha$.
}
\end{figure}

\begin{figure}[hbt]
\begin{center}
\includegraphics[angle=-90,scale=0.5]{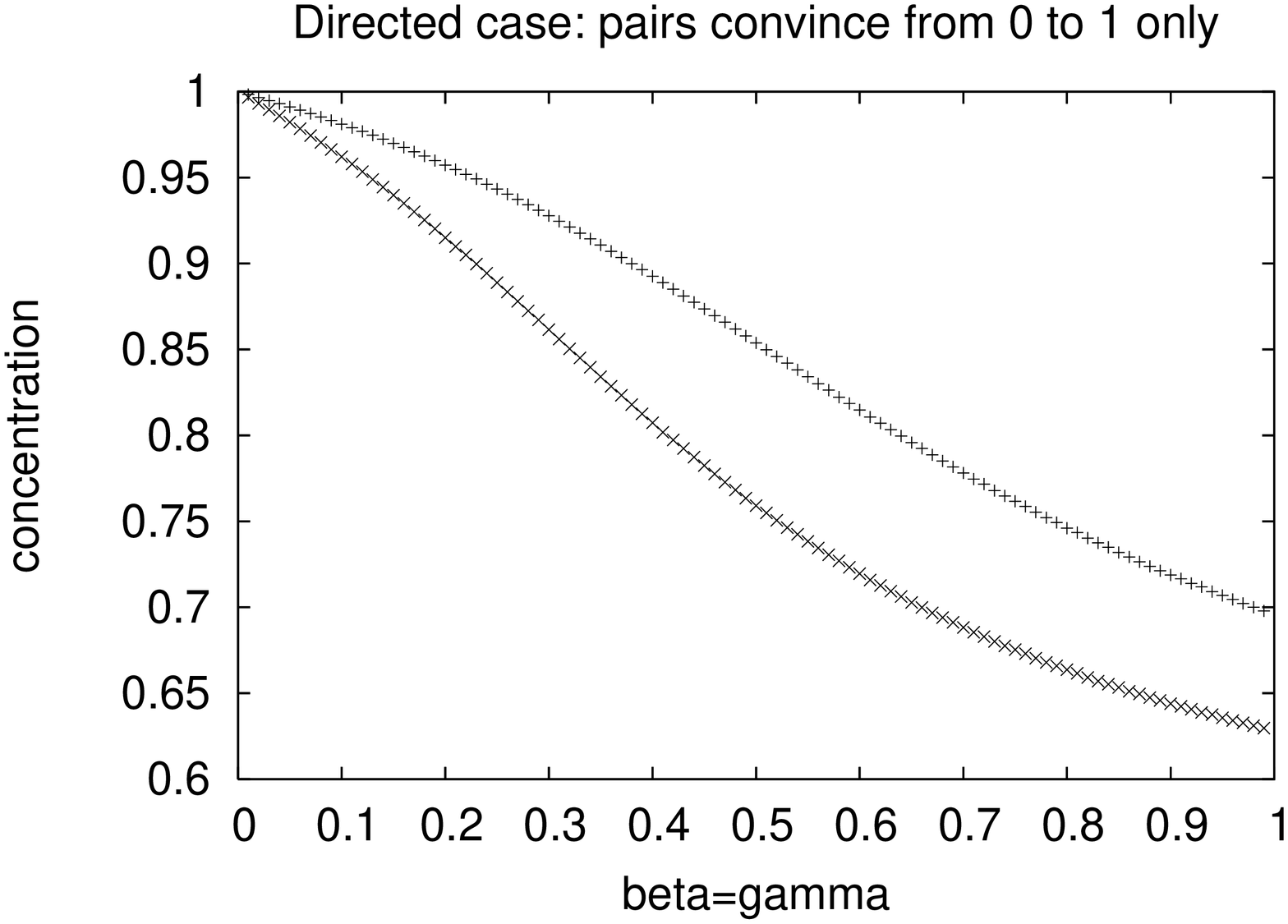}
\end{center}
\caption{No phase transition when the pair-convincing process works only from 
0 to 1. $\alpha = 1$ (+) and 1/2 (x); $L = 201$; $ t = 1000$. Results for
$L = 101$ agree within symbol size with those shown for 201.
}
\end{figure}

A true capital of a country has to coordinate the various regions of that 
country. In this sense, we apply here the suggestion of Nowak (Warszawa) to
modify the Sznajd (Wroc{\l}aw) model of opinion dynamics.

In the Sznajd opinion dynamics model \cite{sznajd,book} on the square lattice
each site (``agent'') can have one of two possible opinions: up (1) or down 
(0). Two neighbouring sites having the same opinion convince their six 
neighbours of that opinion. This model simulates the psychological effect that 
two people in agreement convince better than one person or than two disagreeing 
people \cite{milgram}. A sharp phase transition results for large lattices
if initially a fraction $p$ of the opinions is up randomly: For $p> 1/2$ at 
the end everybody has opinion up, and for $p < 1/2$ everybody ends up with 
opinion down.

A. Nowak at the GIACS summer school ``Applications of Complex Systems to 
Social Sciences'' in September 2006 suggested to generalise this Sznajd model 
by flipping each up opinion down with probability $\gamma$ and each down opinion
up with probability $\beta$, taking into account global effects like advertising
through mass media \cite{advert}. The traditional pair-convincing is 
applied to each of the six neighbours independently with probability $\alpha$.
Obviously, for positive $\beta$ and $\gamma$ no complete consensus is possible
anymore, but we can still search for phase transitions where the fraction of
up opinions jumps as a function of some continuously varying parameter ($\alpha,
\beta, \gamma, p$). 

With $\beta \ne \gamma$ we destroy up-down symmetry; this symmetry can also
be destroyed  by assuming that only up pairs convince down opinions to flip up:
Directed convincing as opposed to the usual undirected convincing. 

The next section brings a mean field approximation, and section 3 our simulation
results, followed by a concluding section 4.

\section{Mean-Field Approximation}

In the symmetric case when $\gamma=\beta$, the appropriate Master equation \cite{gard} 
for the probability $p$ of having up opinion, $S=1$, is 

\begin{equation}
\frac{dp}{dt}=-\beta p +\beta (1-p) + \alpha p^2 (1-p) - \alpha p (1-p)^2
\end{equation}
In this equation, the first term on the r.h.s. is responsible for spontaneous switching from $S=0$ to
$S=1$, the second term - for the opposite, the third - for the switching from 0 to 1 when two neighbours are 1,
and the fourth term - for the switching from 1 to 0 when two neighsours are 0. In the mean field 
approximation, the spatial neighbouring between those neighbours is neglected. Introducing a 
quasi-magnetisation $m\equiv 2p-1$ we get

\begin{equation}
\frac{1}{2} \frac{dm}{dt}=-\beta m + \frac{\alpha}{4}m(1-m^2)
\end{equation}

For the fixed points $m^*$ where $dm/dt=0$, the only relevant is the parameter 
$x\equiv \beta/\alpha$. As the result we get $m^*=0$ or $\pm (1-4x)^{1/2}$. The fixed point is stable 
if the derivative of the r.h.s. of the last equation with respect to $m$ is negative, and it is
not stable if it is positive. Here $m^*=0$ is stable if
$x>1/4$, and the remaining two fixed points exist and are stable if $x<1/4$. This is an example of the supercritical pitchfork
bifurcation \cite{glen}. As we see, the result of the mean field theory is that the phase transition
exists and is continuous. The transition point is at $\beta_c = \alpha /4$. Above this value, the 
spontaneous flipping destroys the correlations between neighbours; those correlations enable the 
Sznajd process and lead to the ordering. For $\beta = \gamma=0$, the result $m^*=0$ or $\pm 1$ agrees with 
former mean-field approach to the Sznajd model \cite{sla}.

\section{Simulations}

Figure 1 shows the final configuration in a $201 \times 201$ square lattice
after 1000 iterations (sweeps through the lattice. random sequential updating).
We use here $p = 0.01$ initially and $\alpha=1$; $\beta = \gamma = 0.15$ in the
upper and 0.20 in the lower part of this figure. Figure 1 shows effects of our 
helical boundary conditions in horizontal direction, since the clusters ending 
at the right border are continued on the left border. We see short-range 
correlations due to the pair-convincing process: The final opinions are not 
distributed randomly in space even though we started with opinions randomly 
distributed on the lattice.

Fig.2 shows more quantitatively the transition between $\beta = \gamma = 0.1$
and 0.2: For low flipping probabilities $\beta, \gamma$ the pair convincing
process dominates and most opinions follow the initial majority. For higher
flipping probabilities near and above 0.2 the flipping probabilities overwhelm 
the pair-convincing process and half of the opinions become up, the other
half down. We see  overlapping curves independent of $L$.  The lower part is an 
expanded plot for the transition region of the upper part. Thus there is a
second-order phase transition at $\beta_c = \gamma_c = 0.13 \pm 0.01$ since for
a jump (first-order transition) the curves would become steeper and steeper 
for larger and larger $L$.

The same transition can also be observed starting from $p=1/2$, Fig.3, instead
of 0.01 as in Fig.2. There, for low flipping probabilities one of the two
opinions randomly has to win over the other, which takes more time; thus only
smaller $L$ are shown in Fig.3. 

In an Ising model, the magnetisation shows a second-order transition, as in 
Fig.3, if the temperature is varied at zero magnetic field, but a first-order
transition (jump) if the magnetic field is varied at constant temperature
below $T_c$. Somewhat analogously, Fig.4 shows this first-order transition
as a function of initial concentration at fixed low $\beta = 0.10$ but not
at higher $\beta = 0.15$.

In all the above figures we had $\alpha=1$; for $\alpha < 1$ the pair convincing
process not always works, and according to Fig.5 $\alpha > 1/2$ at $\beta=\gamma
=0.1$ is needed to preserve the phase transition (initially, $p=0$).

Finally Fig.6 shows the lack of the phase transition if the symmetric 
pair-convincing process is replaced by the directed one: Two neighbouring
up opinions convince the six neighbours; two neighbouring down opinions 
convince nobody. We get the same results whether we start from $p = 0.01$
or 0.99, and whether we use $L=101$ or 201.  Also if up pairs convince
with probability 1 and down pairs convince with probability 1/2, we get 
the same smooth curve whether we start with $p=0.01$ or 0.09.

\section{Conclusion}

With a minor addition ($\gamma$) to the previously simulated Sznajd model with 
advertising \cite{advert} we could see a new second-order phase transition 
in this opinion dynamics. In contrast to \cite{advert} the position of the
transition is independent of $L$. On one side the traditional pair-convincing 
process dominates, on the other side the random opinion flips. This phase
transition is predicted by a mean field theory, but the mean-field position
of the transition is twice as high as simulated for $\alpha=1$. 

We thank A. Nowak for suggesting this work.

\end{document}